\documentclass[submission,copyright,creativecommons]{eptcs}
\providecommand{\event}{TFPIE 2012}
\usepackage{breakurl}
\usepackage{graphicx}

\title{Connecting the Dots: Computer Systems Education\\ using a
  Functional Hardware Description Language}

\author{John T. O'Donnell
  \institute{School of Computing Science\\ University of Glasgow}
  \email{john.odonnell@glasgow.ac.uk}}

\def\titlerunning{Computer Systems Education using a Functional HDL}
\def\authorrunning{J. T. O'Donnell}

\begin{document}

\maketitle

\begin{abstract}
  A functional hardware description language enables students to
  gain a working understanding of computer systems, and to see how
  the levels of abstraction fit together.  By simulating circuits,
  digital design becomes a living topic, like programming, and not
  just a set of inert facts to memorise.  Experiences gained from
  more than 20 years of teaching computer systems via functional
  programming are discussed.
\end{abstract}

\section{Introduction}
\label{sec:introduction}

Since 1991, a variety of courses on computer systems in the Department
(now the School) of Computing Science at the University of Glasgow
have been presented by the author using a functional hardware
description language.  Approximately 35 one-semester courses over 20
years have used these methods.  The language is called Hydra, and is
currently implemented by embedding in Haskell. Development of the
approach to teaching computer systems and research on the Hydra
language have interacted strongly, and each has influenced the other.
A paper discussing how an earlier version of Hydra was used in
teaching computer architecture appeared in 1995
\cite{1995-OD-Hydra-FPLE}.  This paper gives an overview of the
current approach, which has evolved significantly since then, and
discusses the experience that has been gained.

A sign of maturity of functional programming is its quiet
application to practical problem domains, including education,
without too much emphasis on the functional language itself.  The
focus in this paper is on the challenges in understanding computer
systems, and how these challenges can be met successfully.  The
emphasis is on the subject area, not on the language.  A thesis of
this paper, however, is that the aims of teaching computer hardware
are supported very well by a functional hardware description
language, and rather poorly by traditional imperative ones.  The
final result is not a claim that ``functional is better'', but
simply that students gain a deeper insight into computer systems.

Section \ref{sec:rationale} discusses the rationale behind this
approach to teaching computer systems, and Section
\ref{sec:course-topics} summarises the main topics covered.  It is
crucial to be explicit about circuit models, for reasons discussed in
Section \ref{sec:models}.  Section \ref{sec:hydra} briefly describes
the Hydra hardware description language, while Section
\ref{sec:examples} gives some examples.  Section \ref{sec:formal}
discusses opportunities for incorporating formal methods in hardware
design, and Section \ref{sec:exercises} outlines several exercises
that have been used in recent years.  Section \ref{sec:observations}
discusses a number of observations gained through years of experience
teaching computer systems with functional programming, and Section
\ref{sec:conclusion} draws some conclusions.

\section{Rationale}
\label{sec:rationale}

Computer science education has traditionally focused on
programming and software development.  It is common to provide some
courses on computer systems, including an introduction to digital
circuit design and computer architecture, but these courses often
lack the depth found in software engineering.  Standard topics
include an introduction to digital circuits, instruction set
architecture, and basic processor organisation.

A major problem with typical courses on computer hardware is a failure
to ``connect the dots'': various details of each level of abstraction
are covered, but the crucial connections between the levels are
largely ignored.  To understand computer systems properly, it is
necessary to have a \emph{working understanding} of each level of
abstraction, and also to understand the \emph{connections} between the
levels.  To achieve this goal, it is essential to select the topics to
be covered in each level carefully, so the techniques presented in one
level are sufficient for supporting the next level up.

What is meant here by a ``working understanding'' goes beyond just
knowing a collection of facts (the names of the registers, the names
of the techniques, the names of the events that occur).  A working
understanding implies the ability to hand-execute the system, as well
as to design parts or all of it, and to modify it.  A working
understanding provides the foundation for quantitative assessment of
performance.

In contrast, many courses and textbooks
(e.g. \cite{PH1998-CompOrgDesHSInterface}) on computer systems
describe the hardware with schematic diagrams, but do not provide a
working understanding.  This is reflected in many textbook problems
that ask for descriptive answers that can be culled from the text.

Here is an example of what a working understanding means.  There is a
technique in processor design called bypassing, which can speed up a
processor by eliminating pipeline stalls and thus increasing the
amount of parallelism that is achievable.  Many textbooks describe
bypassing, sometimes in considerable detail, but it is easy to
puncture the veneer of a student's understanding by asking some
probing questions: \emph{Here is a fragment of machine code; how many
  clock cycles faster will it run with bypassing?  Modify a processor
  to implement bypassing (i.e. design absolutely every logic gate and
  signal needed to do it).  What performance penalty, if any, is
  introduced by bypassing?}  Unfortunately, there is not room in this
paper to provide a working understanding of bypassing, but it can be
done in a 20-lecture course using a powerful hardware description
language---and Hydra is ideal for this purpose.

Consider a thought experiment.  What if computer programming and
software engineering were taught descriptively?  There would be
explanations of what it's like to write a program, discussions of
how to organise your thinking, and of course a mass of UML
diagrams.  But suppose the students never write a program, and
never hand-execute any code.  The obvious objection is that they
would not be able to do real-world programming after such an
education.  But a further objection is that the students would not
have a working understanding of programming: they would not
understand conceptually what a computer is doing.  (There are
indeed some courses like this, with titles like ``Computing for
Poets'', but they are intended for non-computing students.)

For many computer science students, hardware topics are taught
descriptively, and the students fail to gain a working
understanding.  Two reasons for this are:

\begin{itemize}
\item Computer systems are complex, with many levels of
  abstraction, and each level is a large subject on its own.  There
  isn't enough time to cover all this material.
\item Hardware is seen as an ancillary topic; it's just there to
  run software, which is the important part of computing.
  According to a cliche, \emph{computer science is not about
    computers}, so presumably there is no need to find out how they
  work.
\end{itemize}

\noindent
A response to these problems is:

\begin{itemize}
\item Although systems are complex, it isn't necessary to cover all
  the details of all the levels.  With suitable notation (this is
  where the functional language enters), the essential details can be
  covered clearly and concisely, providing a working understanding.
  In addition, we can achieve a working understanding of how each
  level of abstraction connects to the levels above and below it.
  However, it is necessary to present the right concepts at each
  level.  If we cover precisely the techniques needed to support the
  next level up, then we can explain the connection precisely.  Sadly,
  many courses and textbooks cover a random selection of topics at
  each level, while ignoring the topics that are needed to connect the
  levels.
\item Hardware is just as much about algorithms as software is.  There
  is indeed a distinction between hardware and software, but there are
  also profound similarities.  Students often do not perceive these
  similarities, and this is partly the result of ineffective languages
  for describing the hardware, which unnecessarily obscure the
  algorithmic content of the hardware.
\end{itemize}

We would not have computer science without computers, and surely it
is reasonable for any educated computing professional to have a
good working understanding of how a processor executes programs.
Fortunately, this does not require excessive learning time.

Several courses and supporting textbooks follow a similar philosophy,
using software tools and careful selection of topics to enable the
main ideas in computer systems to be connected with each other.  A
book by Nisan and Schocken \cite{NisanSchocken2005:ElementsCompSys},
supported by software tools, introduces circuits, architecture, and
systems software at an elementary level.  A book by Harris and Harris
\cite{HH:DigDesignCompArch} covers circuit design and processor
architecture at a more advanced level, and uses standard imperative
hardware description languages (SystemVerilog and VHDL) to present a
MIPS processor.

\section{Course topics}
\label{sec:course-topics}

Glasgow has a four-year undergraduate honours degree, as well as
1-year taught postgraduate MSc degrees.  Over the years, the author
has taught courses on computer systems at undergraduate levels 2,
3, and 4, as well as MSc.  Naturally, course structures, approaches
to teaching, and even the degree structures change over time.  It
seems more useful to describe briefly the current courses, and how
functional programming is used currently, without going through the
historical evolution of this approach.

In first year, students have two concurrent full-year courses.  One
introduces programming in Python, and the other covers a number of
topics, including databases, human computer interaction, and
computer systems.  The systems material includes an introduction to
the idea of instructions (using a predecessor to the Sigma16
architecture mentioned below), and an introduction to digital
circuits.  The circuit material introduces logic gates, flip flops,
and basic circuits involving a handful of components.

\subsection{General course on computer systems}
\label{sec:CS2-course-topics}

There is a second year course on computer systems which is required
for all honours students.  This course is taught by the author, and
covers three of the main levels of abstraction in computer systems:
(1) instruction set architecture, the central core level; (2) digital
circuits, which are lower level than the instruction set and whose aim
is to implement the instruction set; and (3) operating systems, at a
higher level of abstraction which requires services provided by the
instruction set.

The first part of the course (4 weeks) begins with the middle level of
abstraction, instruction set architecture (ISA).  The approach is to
cover the Sigma16 architecture \ref{sec:isa-sigma16} in complete
detail.  This is supported by tools in Hydra, including an assembler,
linker, emulator, and GUI.  The aim is to give a comprehensive
explanation of instruction sets and how they support the needs of
programming languages and operating systems (i.e. how the instruction
set level supports the next levels up).  Particular attention is given
to instruction representation and the distinction between machine
language and assembly language.  Programming is introduced by showing
how basic constructs from programming languages are translated to
assembly language, including assignments, if-then and if-then-else,
while loops, for loops, and procedures.  This gives a good opportunity
to give a brief introduction to compilers (what they do, not how they
work) and top-down structured programming.

The second part of the course (3 weeks) covers digital circuits, and
uses a schematic capture software tool to describe simple circuits
(schematic capture uses a graphical tool to draw a circuit diagram and
extract its structure to allow simulation).  The chief emphasis is on
the synchronous model of digital circuits (see Section
\ref{sec:models}).  Circuit design is presented through a sequence of
examples, which are carefully chosen to build up to a circuit called
RTM (register transfer machine) that can execute some aspects of
Sigma16 architecture.  The concepts of cache and virtual memory are
introduced but not implemented.

The third part of the course introduces key concepts in operating
systems and networks, including processes and interrupts, memory
protection, user and system states, address spaces and address
translation.  This material is presented descriptively, as there is
not time to show all the details.  The emphasis is on how higher level
concepts are supported by hardware features; for example, the
operating system concept of processes is implemented using the
hardware technique of interrupts.

The course includes two assessed exercises which provide an
opportunity to work with concrete designs.  These exercises are
supported by scheduled lab sessions with tutors.  In the first
exercise, students write and test an assembly language program in
Sigma16 using the Hydra emulator.  The second exercise covers digital
circuits; the students are given the RTM circuit and perform a number
of experiments with it, and then make an extension to it.

\subsection{Advanced course on computer architecture}
\label{sec:CA4-course-topics}

There is another required systems course for third year students,
which is concerned with distributed programming, concurrency, and
operating systems.

The fourth year is primarily devoted to elective courses that cover
a wide range of computer science subjects in depth.  One of these
is Computer Architecture 4, taught by the author using Hydra.  This
course contains 20 lectures, 10 tutorials, and two assessed
exercises.  The main topics are:
\begin{itemize}
\item A comparative survey of instruction set architectures.  This
  focuses on several styles of commercial architecture, but also
  includes a brief review of Sigma16.
\item A systematic approach to designing synchronous circuits,
  using Hydra for specification and simulation.  Some advanced
  techniques which are needed for processors are covered: ALU
  design, functional units, datapath, and control.
\item Design and operation of a sequential processor; this is a
  digital circuit (called M1) that implements the Sigma16
  instruction set architecture.  The circuit is given in complete
  detail, and the same programs that can run on the Sigma16
  emulator can also be executed by simulating the circuit.
\item Analysis of processor performance, and methods for speeding
  up execution, including cache, pipeline, and superscalar.
\item Introduction to parallel architectures.
\end{itemize}

The core of this course---the central part that establishes the
connection from logic gates and flip flops all the way to a
complete processor circuit---comprises 11 lectures and 5 tutorials.

To put this in perspective, the excellent course by Winkel and
Prosser \cite{WP1986-ArtDigDesign} covers similar material, but 90
lectures over two semesters are required, in addition to a similar
amount of lab time. In the Winkel \& Prosser course, students
actually fabricate a processor specified with schematic diagrams
using MSI components, while the Glasgow course is based on
simulation of functional specifications.  Most courses and
textbooks on computer architecture simply avoid implementing a
processor as a digital circuit, perhaps because it is felt that
there is not enough time.

Probably the most important result from this work is that
connecting the dots, so that students really understand how a
computer works, does not require several hundred hours of class and
lab time: the combination of functional programming and simulation
enables it to be achieved in just half of a 20 lecture course.
This means that \emph{every computer science student could learn
  how computers work.}

\section{Models and synchronous circuits}
\label{sec:models}

It is essential to use appropriate abstract models while working with
computer systems.  Failure to do so can lead to many misconceptions,
and also interferes with the practical use of software tools.

Many textbooks on computer hardware take a low level, bottom up
approach.  A large set of primitive components are presented, and some
discussion is given of how the components behave when connected
together.  This is analogous to very early (1950s) books on
programming, which start by explaining what the statements do, then
just build up bigger and bigger programs.  This bottom up approach to
programming largely died out when structured programming became
popular, but it is still often used in teaching digital circuits.

One disadvantage of a purely bottom up approach is that it requires
covering the design of building block circuits whose purpose will
become clear only much later.  A purely top down approach is also
ineffective, because the higher level circuits seem vague without
understanding something of the underlying technology.  The author's
experience is that a mixed approach works better than pure bottom up
or top down.

The author uses the synchronous model exclusively for teaching, apart
from a short portion of the advanced course on architecture, where the
issues of metastability and synchronisation are covered.  This model
assumes that signal values are restricted to 0 or 1; the combinational
circuits contain no feedback; that all flip flops receive a clock tick
at the same point in time; that input signals to the circuit become
valid at a clock tick (i.e. the value of an input signal cannot change
at a random point in time); and that the clock runs slowly enough to
ensure that all signals are valid before a clock tick occurs.

It is relatively straightforward to design a circuit so that the
restrictions for the synchronous model are satisfied.  The benefit is
that the circuit behaves like a state machine, where the state is held
in the flip flops and the state transition function is well defined by
the combinational logic gates.

The synchronous model is not just used implicitly; it is a central
topic that is covered in detail in both the general course on systems
and again in the advanced course on computer architecture.
Asynchronous circuits are also important, of course, but it is best to
cover them and all their attendant complications only after
synchronous design has been mastered.

A particularly pernicious situation arises when the synchronous
circuit model is not covered explicitly, and the approach to design is
bottom up (learn the components, then just connect them together).
Although this approach may sound silly, it is in fact very common, is
used in many textbooks, and was in use at Glasgow for the required
level 2 course before the author took it over.  The most important
disadvantage is that the central concepts of state machines are
weakened.  There are also severe practical problems: the synchronous
model avoids race conditions, but without it the presence of race
conditions comes down to luck.

If you design a circuit without following the synchronous model, just
how much harder does it become to predict and understand the
behaviour?  To answer this question, consider the circuit \textit{ x
  \textbf{where} x = inv x}.  The circuit could hardly be simpler: it
consists of an inverter whose input is connected to its output.  This
circuit is not synchronous, because it contains a feedback loop in
combinational logic.  The behaviour is well understood, but it took a
number of research papers to analyse it.  In the computer architecture
course, the circuit is presented, and the students are asked to guess
its behaviour.  To date, none of the author's students has ever
guessed correctly.

The point of this short diversion is not to understand a rather
useless circuit, but to underscore the crucial role that the
synchronous model plays in practical design.  To solve real problems,
circuits are likely to require thousands or even hundreds of thousands
of components.  This is routine with the synchronous model, but
without the model a circuit with just one component becomes
intractable.

Notwithstanding the above, asynchronous circuits are becoming
increasingly important.  However, the usual approach is to use
asynchronous signalling between building blocks that are purely
synchronous but that have independent clocks.  The signals that cross
clock domains must pass through a synchroniser, and there is a
non-zero risk that the receiving circuit may become metastable.  These
issues are fascinating and important in modern large scale designs, so
they are introduced in the computer architecture course.

Models are also necessary for predicting and improving circuit
performance, yet they are commonly ignored.  A good example is the use
of Karnaugh maps.

During the 1950s, computer circuits were constructed from vacuum
tubes, and the cost of a machine depended mostly on the number of
tubes.  Consequently, a chief aim of circuit design was to minimise
the number of tubes.  To this end, a technique called Karnaugh maps
was developed around 1950 to ``minimise'' circuits for Boolean
functions by reducing the number of logic gates.  Karnaugh maps are
still a major topic in many introductory books and courses on digital
circuits.

Modern hardware is implemented on integrated circuits, and the cost of
a circuit bears little relation to the number of logic gates.  Indeed,
many circuits are not constructed from logic gates at all, but use
instead a technique called steering logic that uses transistors
directly.  The \emph{area} of a circuit is far more important than the
number of logic gates, and the best way to minimise the area is to use
a regular geometric layout---which may actually increase the number of
gates.

Karnaugh maps have their merits---they are interesting, they provide a
good example of a simple circuit transformation, and they are useful
for programmable logic arrays such as FPGAs---but surely it is not
worth spending a large fraction of the limited time available in a
computer systems course on an optimisation technique that became
largely obsolete before the students' parents were born.

Rather than discussing Karnaugh maps, it is better to present the idea
of \emph{cost models}, to give several examples of cost models, and to
show how to transform a circuit to lower its cost according to a
suitable cost model.

\section{Functional hardware description with Hydra}
\label{sec:hydra}

The computer systems courses described here use Hydra
\cite{2002-OD-PDSECA-Hydra}, a functional hardware description
language that is embedded in Haskell.  Hydra supports many, though
not all, aspects of computer hardware, including digital circuits,
design methodology, processor organisation, and machine language
programming.  Several levels of abstraction are covered, including
logic gates, combinational and sequential circuits, register
transfer level, datapath and control, synthesis of control circuits
from control algorithms, and machine language.

There are several functional hardware description languages that are
similar to Hydra.  Lava \cite{BCS1998:lava} is similar to the 1987
version of Hydra which uses pointer equality (``observable sharing'')
to determine a circuit's structure.  However, this technique is impure
and has several unfortunate impacts on equational reasoning; it was
abandoned around 1990 in Hydra, which uses program transformations to
generate circuit netlists.

Hydra treats circuits as functions.  A circuit specification is a
function from input signals to output signals.  This can obviously
be done for combinational circuits (logic gates), but the
simplicity and power of Hydra stem from the fact that all
circuits---including sequential ones---can be modelled as pure
functions.  The insight that makes this possible was discovered by
Steven D. Johnson \cite{J84-app-prog-dig-design}, and is based on
using streams to record the entire history of a signal value as a
single denotational value.

A Hydra specification combines the structure and behaviour of a
circuit in a single function definition.  The system provides a
number of alternative semantics, which are selected using type
classes.  By choosing a suitable signal representation, a
specification can be executed, analysed, or a netlist generated,
all by simply executing the specification.

The Hydra software consists of a library, an executable application
with a GUI, and a set of examples.  The library provides facilities
for specifying and generating circuits, as well as extensive tools
for performing simulations.  The examples include a variety of
useful circuits, as well as a complete computer architecture called
Sigma16.  The GUI provides an easy interface to using the tools.

The Sigma16 system is an extended example for Hydra, and is used as
an example of basic techniques in computer systems for the courses.
There have been extensive discussions in the literature about the
tradeoffs between using real architectures or synthetic ones (like
Sigma16), and also debates about using emulation vs. execution on
real hardware.  A synthetic architecture like Sigma16 cuts out a
great deal of irrelevant detail, allowing more time to be spent on
the important issues.

There is an assembler for Sigma16, a loader, and emulator, and a
GUI that makes it easy to use these tools.  Figure \ref
{fig:Sigma16-processor-gui} shows the GUI while the emulated
processor is running an example program.  There is a library of
assembly language example programs for Sigma16.

There is also a complete digital circuit that implements the
Sigma16 instruction set architecture.  The circuit comprises logic
gates and flip flops, and a separate memory component.  The circuit
is complete: absolutely every component and wire needed for the
computer is present; thus a full understanding of the processor can
be obtained by studying the complete circuit specification.  And
this is not too hard: the complete circuit specification is about
500 lines of code, many of which are comments.

The student can write a program in assembly language, and translate
it to machine language using the assembler (or by hand).  The
machine language program can be executed in two ways: using the
emulator, and by simulating the digital circuit with the program as
input.  These tools are very effective at connecting the dots
between the levels of digital circuits and computer architectures.

All of Hydra is implemented in Haskell, including the library, the
assembler and emulator, and the GUI.  The library uses Template
Haskell, and requires the ghc implementation of Haskell.  The system
can be used either with ghci (interpreter) or ghc (compiler).  The
ghci interpreter is fast enough to execute a machine language program
running on the Sigma16 digital circuit as it executes a machine
language program.

\section{Examples of functional circuit specification}
\label{sec:examples}

Hydra is used for teaching instruction set architecture in a
required course for second year computing science students, and for
teaching circuit design and computer architecture in an elective
course for fourth year students.

Some snippets of materials used in these courses are given below,
although they are minimised for this extended abstract.

\subsection{Instruction set architecture}
\label{sec:isa-sigma16}

Both the elementary and advanced course use Sigma16 as an example
architecture.  This is a RISC style architecture modelled on the
MIPS.  It has 16-bit words, and 16 general registers.  There are
two instruction formats: RRR (where the three operands are all in
registers), and RX (where one operand is in a register and the
other is specified as a memory address with an index register).

Some of the RRR instructions are:

{\footnotesize
\begin{verbatim}
op mnemonic    operands    action
0    add       R1,R2,R3    R1 := R2+R3
1    sub       R1,R2,R3    R1 := R2-R3
  ...
4    cmplt     R1,R2,R3    R1 := R2<R3
5    cmpeq     R1,R2,R3    R1 := R2=R3
6    cmpgt     R1,R2,R3    R1 := R2>R3
  ...
d    trap      R1,R2,R3    trap interrupt
e             (expand to XX format)
f             (expand to RX format)
\end{verbatim}
}

The RX instructions use an expanding opcode (and this useful
architectural technique is explained fully in the courses---a
working understanding is interesting and useful for anyone who
needs to learn about compilers).  Some of the RX instructions are:

{\footnotesize
\begin{verbatim}
op sb  mnemonic  operands   action
f   0  lea       Rd,x[Ra]   Rd := x+Ra
f   1  load      Rd,x[Ra]   Rd := mem[x+Ra]
f   2  store     Rd,x[Ra]   mem[x+Ra] := Rd
  ...
f   5  jumpt     Rd,x[Ra]   if Rd/=0 then pc := x+Ra
f   6  jal       Rd,x[Ra]   Rd := pc, pc := x+Ra
\end{verbatim}
}

\begin{figure}[t]
  \begin{center}
    \includegraphics[angle=0,scale=0.39]{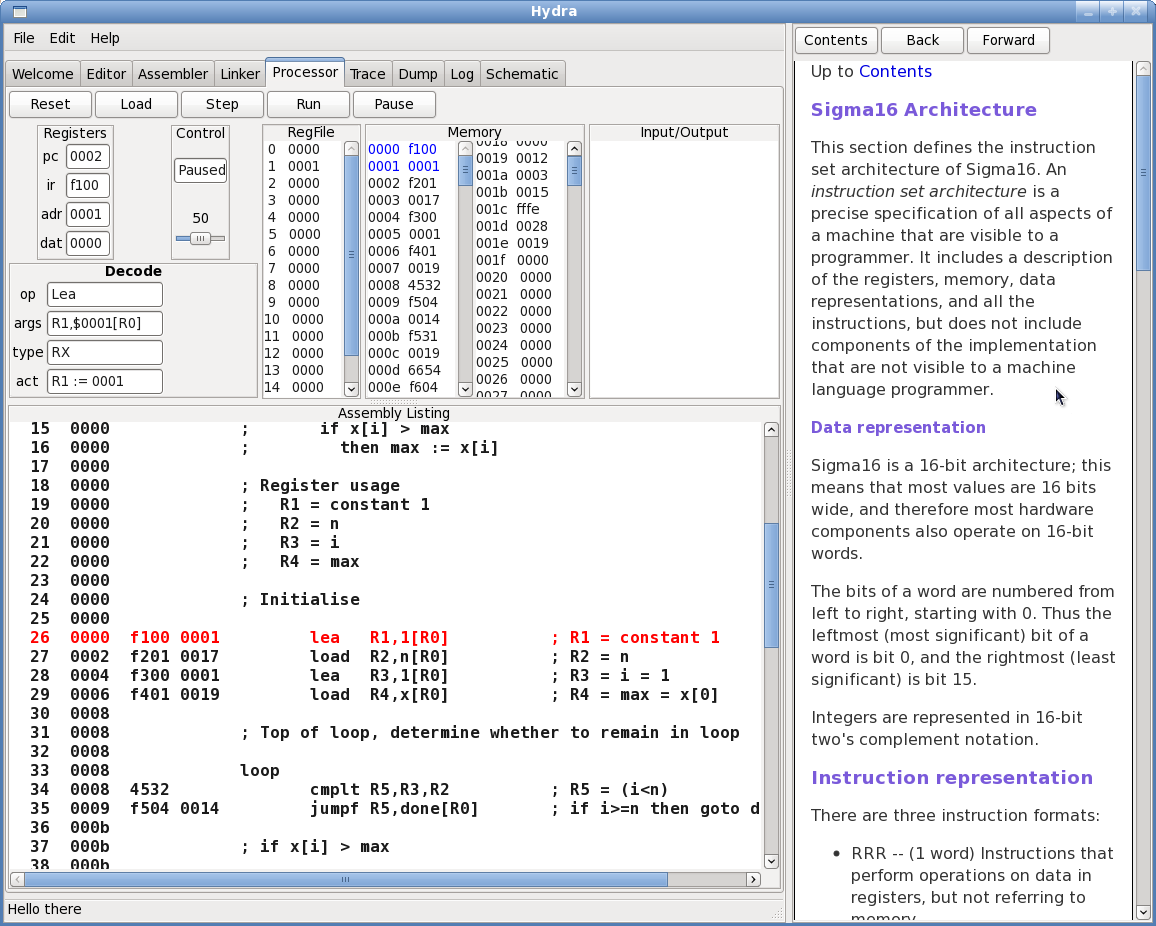}
  \end{center}
  \caption{Hydra GUI.  The \textbf{Processor} tab is selected, and
    the machine is running the ArrayMax program.}
  \label{fig:Sigma16-processor-gui}
\end{figure}

Many of the instructions are illustrated in the program ArrayMax,
which demonstrates basic arithmetic, arrays, comparisons and
conditional jumps, and loops.  Figure
\ref{fig:Sigma16-processor-gui} shows the GUI for the emulator as
it is executing ArrayMax.

\subsection{Digital circuits}

Hydra treats a digital circuit as a function from input signals to
output signals.  This approach is obvious for combinational circuits,
but not for sequential circuits that contain flip flops with state.  A
key insight is that all circuits, including sequential ones, are pure
functions if we model a signal as a stream of values through time,
rather than as a value at a point in time.

Since circuits are functions, it becomes easy to specify them in a
functional language.  Here is the specification of a multiplexer, with
control input \texttt{c} and data inputs \texttt{x} and \texttt{y}:

\begin{verbatim}
mux1 :: Signal a => a -> a -> a -> a
mux1 c x y = or2 (and2 (inv c) x)
                 (and2 c y)
\end{verbatim}

The type specification (which is optional) says that the circuit has
three inputs and one output, and each is a bit signal.  The defining
equation uses the identifiers c, x, and y as local names for the input
signals.  It gives the value of the output signal as a function of the
inputs.

A register circuit takes a control input ld, and a data input x.  It
contains an internal state, which is always available on the output.
At a clock tick, the register retains its previous state if ld=0, and
replaces the state with x if ld=1.

\begin{verbatim}
reg1 :: Clocked a => a -> a -> a
reg1 ld x = s
  where s = dff (mux1 ld s x)
\end{verbatim}

The type indicates that reg1 is a synchronous circuit, and its inputs
and outputs must have a signal type that knows about clocking.

The examples above show that basic circuits are quite straightforward
to specify.  These are executable specifications, and by selecting
different types to represent the signals, alternative semantics can be
selected.

Larger circuits involve large numbers of signals, and it is unwieldy
to treat them all as singleton bits.  Therefore Hydra allows signals
to be grouped together as words and tuples, using the Haskell types
for lists and tuples.

Hydra also supports circuits with size parameters, combinators, and
generators for specialised kinds of circuit.  A circuit generator is a
higher order function that generates a circuit, which is itself a
first order function.  A simple example of a circuit generator is a
map function that takes a building block circuit that takes bit inputs
and outputs, and replicates the building block to produce a circuit
that operates on words rather than bits.  A more complex circuit
generator takes a control algorithm, expressed as a state machine
algorithm, and generates a control circuit.

\subsection{Datapath and control}

An important concept in designing complex circuits is to partition the
design into a \emph{datapath} and \emph{control}.  The datapath
contains the registers and the circuits that perform calculations.
The datapath for Sigma16 includes the following equations, which
define a set of registers and the output of the ALU:

\begin{verbatim}
  (a,b) = regfile n k ctl_rf_ld ir_d rf_sa rf_sb p
  ir = reg n ctl_ir_ld memdat
  pc = reg n ctl_pc_ld r
  adr = reg n ctl_ad_ld q
  (ovfl,r) = alu n ctl_alu_op x y
\end{verbatim}

One of the most interesting aspects of a datapath is that it provides
a set of alternative potential operations that can be performed, and
these must be supported by multiplexers with corresponding control
signals.  For example, the first data input to the ALU is called
\texttt{x}, and sometimes this should be the value of \texttt{a} (a
readout from the register file) and sometimes it should be the
\texttt{pc} (in order to increment the \texttt{pc}).  To support this,
we define \texttt{x} to be the output of a multiplexer that selects
between \texttt{a} and \texttt{pc}, and introduce a control signal to
determine which value to use.  The datapath contains a number of
similar equations.

\begin{verbatim}
  x  = mux1w ctl_x_pc a pc        -- alu input 1
\end{verbatim}

While the datapath provides potential operations, the control uses
a collection of control signals to determine which operations
should actually take place during the current clock cycle.
There are about 20 control signals in the basic Sigma16 M1 circuit.
A few of them are:

{\footnotesize
\begin{verbatim}
  ctl_rf_ld    Load  register file (if 0, remain  unchanged)
  ctl_x_pc     Transmit pc on x (if 0, transmit reg[sa])
  ctl_y_ad     Transmit ad on y (if 0, transmit reg[sb])
  ctl_rf_alu   Input to register file is ALU output (if 0, use m)
\end{verbatim}
}

\subsection{Control algorithm}

Textbooks on circuit design often treat control as just another
circuit, albeit a large and complex one.  There are more
sophisticated design methodologies that use Algorithmic State
Machines (i.e. flowcharts) to describe the control.  One of our
conclusions, however, is that it is best to treat control as an
algorithm that is expressed in a language, and then to develop
systematic methods for synthesising the control circuit.

The control algorithm for Sigma16 is a state machine written in an
imperative style.  The top of the main loop repeatedly fetches the
next instruction and decodes it:

{\footnotesize
\begin{verbatim}
repeat forever
    ir := mem[pc], pc++;
    case ir_op of
      ...
\end{verbatim}
}

There is a separate case for each instruction; the case for the
load instruction (1) fetches the second word of the instruction
(the displacement), (2) calculates the effective address, and (3)
fetches the data from memory and loads it into the destination
register:

{\footnotesize
\begin{verbatim}
  load:
       ad := mem[pc], pc++;
       ad := reg[ir_sa] + ad
       reg[ir_d] := mem[ad]
\end{verbatim}
}

The control algorithm is a finite state automaton.  Alternatively, one
can think of the control as a program running on the datapath, which
can be thought of as a programming language.  This view is natural,
because the datapath provides a set of primitive capabilities (e.g.
``send the value in the \texttt{pc} to the ALU'', ``tell the ALU to
increment its input'', and ``copy the output from the ALU back to the
\texttt{pc}'').  By combining and sequencing such primitive
operations, the control algorithm causes the datapath to perform
useful computations.  In order to make the datapath perform the
operations specified by the control algorithm, we need to figure out
which control signals must be asserted.  In general, this can be
assisted by software tools, but for a student just learning how a
processor works, it's best to work this out by hand, at least for
several instructions.

The following excerpt shows the control algorithm for implementing a
load instruction.  There are three states, so the algorithm executes
in three clock cycles.  The first state fetches the address field of
the instruction; the second state calculates the effective address;
the third state loads the data into the destination register.

{\footnotesize
\begin{verbatim}
  st_load0:
    ad := mem[pc], pc++;
       Assert [ctl_ma_pc, ctl_adr_ld, ctl_x_pc,
               ctl_alu_abcd=1100, ctl_pc_ld]
  st_load1:
    ad := reg[ir_sa] + ad
       Assert [set ctl_y_ad, ctl_alu_abcd=0000,
               set ctl_adr_ld]
  st_load2:
    reg[ir_d] := mem[ad]
       Assert [ctl_rf_ld]
\end{verbatim}
}

\subsection{Running programs on the circuit}

The basic circuit (version M1) for the Sigma16 architecture can
execute machine language programs, simply by simulating the
circuit.  However, the circuit contains many input and output
signals, and thousands of internal signals.  It would be impossible
to figure out what is going on by looking at a bit-level simulator.

\begin{figure}[t]
\hrule
\vspace{0.6em}
{\footnotesize
\begin{verbatim}
Clock cycle 67
Computer system inputs
         reset=0 dma=0 dma_a=0000 dma_d=0000
ctl_start = 1

Control state
   st_instr_fet = 0  st_dispatch = 0       st_add = 0       st_sub = 0
        st_mul0 = 0     st_cmplt = 0     st_cmpeq = 0     st_cmpgt = 0
       st_trap0 = 0      st_lea0 = 0      st_lea1 = 0     st_load0 = 0
       st_load1 = 0     st_load2 = 0    st_store0 = 0    st_store1 = 0
      st_store2 = 0     st_jump0 = 0     st_jump1 = 0    st_jumpf0 = 0
      st_jumpf1 = 1    st_jumpt0 = 0    st_jumpt1 = 0      st_jal0 = 0
        st_jal1 = 0

Control signals
   ctl_alu_a  = 0 ctl_alu_b  = 0 ctl_alu_c  = 0 ctl_alu_d  = 0
   ctl_rf_ld  = 0 ctl_rf_pc  = 0 ctl_rf_alu = 0 ctl_rf_sd  = 0
   ctl_ir_ld  = 0 ctl_pc_ld  = 1 ctl_ad_ld  = 0 ctl_ad_alu = 0
   ctl_ma_pc  = 0 ctl_x_pc   = 0 ctl_y_ad   = 1 ctl_sto    = 0

Datapath
    ir = f604  pc = 0010  ad = 0011   a = 0000   b = 0012   r = 0011
     x = 0000   y = 0011   p = 0331  ma = 0011  md = 0000 cnd = 0

Memory
   ctl_sto = 0      m_sto = 0
     m_addr = 0011  m_real_addr = 11  m_data = 0000  m_out =0331

Fetched displacement = 0011
jumpf instruction jumped
************************************************************************
Executed instruction:  jumpf  R6,0011[R0]   effective address = 0011
jumped to 0011 in cycle 67
Processor state:    pc = 0010  ir = f604  ad = 0011
************************************************************************
\end{verbatim}
}
\hrule
\caption{Simulation output.}
\label{fig:simulation-output}
\end{figure}

Hydra contains a sublanguage for expressing simulation drivers
(also called testbenches).  This is a piece of software that
accepts input from the user in a readable format, converts it to
the bit signal representations, connects the input signals to the
circuit, executes the circuit (thereby simulating it), monitors the
circuit's output signals, converts their values to a readable form,
and prints that out.  Figure \ref{fig:simulation-output} shows the
output from the simulation driver, as the circuit is on clock cycle
67 while executing the ArrayMax program.

Sometimes it is hard to see what is going on, even looking at the
values of registers and signals.  Therefore, the simulation driver
language maintains a state and provides tools that allow the driver
to observe signal values and record partial information as it goes.
The simulation driver for Sigma16-M1 watches the output signals
from the circuit, collects information, and uses that to print an
informative message when a major event (such as the execution of an
instruction) occurs.  The last few lines of the figure show a
message indicating that a jumpf instruction has just executed.

The M1 circuit for Sigma16 takes the simplest approach to solve all
the problems needed to execute computer programs.  It takes several
hours to understand, but average students in the computer
architecture course really do understand it.  Furthermore, they
develop a working understanding, and demonstrate this in exercises
that involve modifications (adding new instructions, implementing
interrupts, etc.).

There are also circuits that introduce more advanced techniques,
including pipelining and superscalar execution.  And, with these
powerful design techniques, it is quite straightforward to
implement different instruction set architectures.

\section{Formal methods}
\label{sec:formal}

Formal methods have been more successful in digital hardware design
than in software design.  One reason for this is that it's quick and
easy to recompile a program: just type \texttt{make} and wait a
minute, but it's slow and costly to refabricate a circuit: rebuild the
masks, send to the foundry, and wait a month.  The lazy approach that
works for programming fails for hardware design.  Another reason is
that it is more costly and damaging for a hardware manufacturer to
ship chips that don't work than for a software vendor to ship software
that doesn't work.

The most popular formal method is probably model checking.  A pure
functional hardware description language also makes it possible to
use equational reasoning, which can be used to prove correctness,
perform correctness-preserving transformations, and even to derive
circuits from specification by calculation.

During the last two years, formal methods have not been used in the
fourth-year course on computer architecture, but they were used
successfully in the past.  (The problem is that there are only 20
lectures available, and far too much material to cover everything.)

A good application of formal methods is in the derivation of a
logarithmic time binary addition circuit \cite{2004-ODR-Adder-JFP}.
This is an excellent example because it yields a very subtle
circuit which is quite hard to understand, yet is essential for a
fast processor.  Indeed, the speedup we obtain by replacing a
ripple carry adder by a log time adder is larger than the speedup
from larger word sizes, or cache memory, or pipelining, or
retiming, or superscalar execution---it is probably the most
effective single optimisation available in processor design.

The solution requires first deriving the parallel scan algorithm,
followed by a sequence of transformations to a ripple carry adder.
Experience shows that a minimum of three lectures are needed for
this, and not all students are able to follow the details.

There are many alternative problems that could be used to
illustrate formal methods in less time, but they do not make such a
convincing case for the power of mathematics.

A limitation of the adder derivation is that this is a
combinational circuit.  Another direction is to prove correctness
of sequential circuits, especially to prove that a circuit
correctly implements an instruction set architecture.  However,
this requires a significant amount of notational machinery
(techniques and lemmas) in order to handle the state.

\section{Exercises}
\label{sec:exercises}

A central aim of our courses is to help the students to develop a
working understanding.  Exercises that involve implementation, as
well as observing the implementation running in simulation, are a
crucial component.  Several exercises from recent years are
outlined below, including examples from both the required
second-year course on computer systems and the elective fourth-year
course on computer architecture.  The exercises range from quite
elementary to fairly advanced.  All of these exercises have been
solved largely correctly by a majority of students, although there
are various glitches and infelicities in many of the solutions.

\subsection{Assembly language programming: insertion sort}

This exercise is used in the required course on computer systems
for second year students.

\paragraph{Problem.} Translate the insertion sort algorithm, which
is given as high level pseudocode, into Sigma16 assembly language.
Test it on an array of input data using the assembler and emulator.
Check that the program is working correctly during execution, and
verify that after the program terminates the array is sorted
correctly.

\paragraph{Comments.}  The lectures emphasise the relationship between
high level language constructs (like the statements in the
pseudo-code) and instructions.  Students are shown how to compile an
algorithm to assembly language by hand, and strongly encouraged to do
this rather than writing the assembly language code directly.  This
problem requires ability to handle memory, registers, arrays, index
arithmetic, looping, and some simple logical constructs.  A systematic
commenting style is used consistently in lectures and tutorials, and
the students are told that the marking will include assessment of
comments.  Figure \ref{fig:Sigma16-processor-gui} shows a snapshot of
the emulator while executing the insertion sort program.

This exercise has been refined over several years, using a number
of different algorithms, in order to get a good balance between
simplicity and richness of insight.  The aim is to give a good
understanding of how the instructions work, how they relate to the
algorithm, and in general what the machine is doing, but the aim is
also to achieve this without needing to write large amounts of
boilerplate code.

Many students are completely baffled by this problem when they
start. but they receive good support from tutors.  The vast
majority of students not only succeed in getting this to work, but
they also do reasonably well on an unseen assembly language
programming problem of similar complexity on the final examination.

\subsection{Basic state machines: traffic light controller}

\paragraph{Problem.}
Design a traffic light controller as a digital circuit.  There are
two versions.  The first version has one input, a pushbutton bit
called reset, which is pushed once to start the circuit.  There are
three output bits corresponding to green, amber, and red, which run
through a fixed sequence: green, green, green, amber, red, red,
red, red, amber, and so on.  The second version models a pedestrian
crossing, with a walk request input button; the system controls
walk and don't walk lights as well as the coloured lights for
traffic.  This version of the circuit should exhibit reasonable
behaviour even if the walk request button is pressed frequently.
The circuit also maintains a count of walk requests, which could be
used by traffic engineers.

\paragraph{Comments.}
These circuits are very simple, although they do illustrate some
important basic design techniques.  The main point of the exercise
is to get the students to learn how to write a correct
specification in Hydra and run it.  It is better to resolve
problems with notation on an easy problem like this, than to defer
them to a problem where the digital design is challenging.

\subsection{ALU}

\paragraph{Problem.} Design an arithmetic-logic unit (ALU) that
performs integer arithmetic on 8-bit signed integers represented in
two's complement notation.  The interface to the circuit is
specified with a Hydra (Haskell) type declaration.  The circuit
inputs are a two-bit opcode \texttt{op :: (a,a)}, and two 8-bit
words \texttt{x} and \texttt{y}, which are supplied as a group
\texttt{xy :: [(a,a)]} in bit slice format.  The outputs are
\texttt{(ofl, r) :: (a, [a])}, where \texttt{ofl} indicates
overflow, and \texttt{r} is the 8-bit result word.  The value of
the output \texttt{r} depends on the value of the control input
\texttt{op}, as follows:
\begin{center}
\begin{tabular}{|c|c|}
\hline
  \textit{op} & $r$ \\
\hline
  (0,0) & $x+y$ \\
  (0,1) & $x-y$ \\
  (1,0) & $y+1$ \\
  (1,1) & $-y$ \\
\hline
\end{tabular}
\end{center}

\paragraph{Comments.}
This is a combinational design, with no circuit state such as flip
flops.  Several interesting techniques need to be combined to
solve the problem.  The best approach is to use a single binary
word adder, and to make the adder perform all the required
operations by preprocessing the inputs and postprocessing the
outputs.  The mscanr higher order function gives a simple, elegant,
and general definition of the adder.  The bit slice organisation is
useful for incorporating the ALU into larger circuits, and also
provides useful experience with types and patterns.

\subsection{Pipelined integer vector multiplier}

\paragraph{Problem.}
Design a circuit that inputs a pair of $k$-bit binary integers on
every clock cycle.  After $k$ cycles, the circuit outputs the $2
\times k$-bit result.  The circuit needs to be pipelined, as a new
pair of integers are read in \emph{every cycle}.

\paragraph{Comments.}
The students have already been shown a sequential multiplier
functional unit based on the ``shift and add'' algorithm.  The
essence of the problem is to transform a sequential circuit that
uses registers for in-place state into a pipelined circuit that
unfolds the iteration into a sequence of states.

\subsection{Loadxi instruction}

\paragraph{Problem.} Add a new instruction to the Sigma16
architecture, as defined below.  Modify the datapath and control,
as needed, in order to implement the new instruction in the M1
circuit.  Modify the test bench so the operation of the instruction
can be observed, and demonstrate the execution of the instruction
using a machine language test program.

The new instruction is \emph{load with automatic index increment}
(the \texttt{loadxi} instruction). Its format is RX: there are two
words; the first word has a 4-bit opcode f, a 4-bit destination
register (the d field), a 4-bit index register (the sa field), and
a 4-bit RX opcode of 7 (the sb field).  As with all RX
instructions, the second word is a 16-bit constant called the
displacement. In assembly language the instruction is written, for
example, as loadxi R1,\$12ab[R2].

The effect of executing the instruction is to perform a load, and
also to increment the index register automatically. The effective
address is calculated using the old value of R2 (i.e. the value
before it was incremented.)  Thus the instruction loadxi
R1,\$12ab[R2] performs R1 := mem[12ab+R2], R2 := R2+1.

\paragraph{Comments.}  This exercise requires changes to about a
dozen lines of code (including comments).  However, it is necessary
to understand how the datapath, control algorithm, and control
circuit work in order to make those changes.  The assignment
handout asks for a status report that includes an explanation of
how the instruction was implemented, as well as a machine language
program that demonstrates the instruction and simulation output
showing that the instruction works correctly.  A large majority of
the students succeeded, while some others outlined what needs to be
done but didn't complete the modifications.

\subsection{The multiply instruction}

\paragraph{Problem.} The version of Sigma16 that is provided to the
students has a Multiply instruction, but it acts as a ``nop'': it
does nothing at all.  The students have also been given a
standalone multiplier functional unit circuit, as a simple example
of sequential design.  The problem is to make the multiply
instruction work correctly, and to demonstrate its operation by
simulating the processor as it executes a suitable test program.

\paragraph{Comment.}
This problem is about processor design, not about multiplication,
since the multiplier circuit has been provided.  The main
complication is that the multiplier circuit takes a variable amount
of time, depending on the values of the inputs, and the processor
control needs to take account of this.  Changes are required to the
datapath, control algorithm, and control circuit.  In addition,
changes should be made to the testbench definition, so the
operation of the multiplier can be observed as the circuit
operates.

\section{Experience and observations}
\label{sec:observations}

The observations in this section are based on the author's own
experiences.  It would be interesting to compare them with the
experiences of other lecturers at other universities in other
countries.

For the most part, students have reacted enthusiastically to the
approach presented here.  Many of them find it enlightening and fun to
see a digital circuit running computer programs, and they feel they
really understand what is happening when they modify the circuit to
add a new instruction.  Nearly all the students who put in a
reasonable effort are able to learn the language, understand the
circuit examples, and carry out the exercises successfully.

\subsection{Systems issues}

\paragraph{Connecting the dots.}
The deepest benefits from learning about circuits and computer
architecture come from connecting the concepts, not just learning
them in isolation.  This means, for example, using the material on
digital circuits to implement an instruction set architecture, and
using the material on instruction sets to implement core operating
system facilities.  These connections are more valuable than the
specific details at any one level of abstraction.

\paragraph{Executing circuits and programs.}
A great benefit from using a hardware description language is that
students can watch circuits operate by simulating them, and can
execute programs by emulating them or running them on a simulated
circuit.  Simulation helps to get a working understanding of what a
computer is doing, and this understanding is far deeper than what
can be attained by a vague descriptive approach.

\paragraph{Choice of topics.}
Computer systems contain many levels of abstraction, and there is an
enormous amount of interesting material at each one of them.  A tiny
subset of the material at each level must be selected.  Many textbooks
spend too much time covering lots of ancillary details, preventing
them from getting to the most interesting topics (the ones that are
needed to connect the dots).  For example, it is common for textbooks
on digital design to present a wide range of SSI and MSI components,
and a variety of different kinds of flip flop, yet most of those
components became obsolete decades before our students were born.  It
is much better to skip most of the components, and most of the
optimisation techniques, and go straight to the essential components
and methods needed to see how realistic circuits work, and to develop
those circuits toward a processor.

The choice of topic depends heavily on the time available.  Within
20 lectures, it is quite realistic to explain digital components,
circuits, and move up to a complete processor circuit, all within
12 lectures, leaving time for advanced topics like pipelining and
superscalar, to fit within a 20 lecture course.  We have an
existence proof that this is possible!  But such an ambitious goal
requires ruthless care in selection of topics.  If more time is
available, there is naturally a wealth of additional material to
enrich the course.

\paragraph{Simulation or real hardware?}
An endless debate in computer systems education is whether to have
students run exercises on real hardware, or on simulated (or
emulated) hardware.  This debate arises at all levels, including
circuit design and machine language programming.  There are many
advantages of simulation and emulation:
\begin{itemize}
\item A better environment for tracing and debugging is available.
\item A variety of alternatives (e.g. variations on instruction set
  architectures) can be provided, with little overhead.
\item The particular systems studied (machine languages, digital
  circuits) can be designed specifically for the intended purpose,
  and are not encumbered with all the irrelevant complexity that
  comes with ``real'' systems.
\item Aspects of the system that are not central can be glossed
  over, while they may require great complexity on real hardware.
  For example, you can't run a program on  a digital circuit
  without getting the initialisation right, and the initialisation
  can easily be more complicated than all the rest of the system.
  With a simulator, the initialisation can be performed \emph{deus
    ex machina}.
\item Glossing over the minor details leaves more time for the
  essential ideas.  Indeed, it is unusual to have a course that
  starts with circuit design and attains a complete processor
  circuit within 12 lectures, and that would be impossible using
  real hardware.
\end{itemize}
The advantages claimed for using real hardware are unconvincing:
\begin{itemize}
\item ``It motivates students to learn about real products, rather
  than systems that abstract away irrelevant details.''  This claim
  sounds hollow when students are faced with the details of
  incrementing the pc register on a Pentium, or bootstrapping a
  loader.
\item ``Students appreciate the levels of abstraction in computer
  systems better when they know that they are seeing real software
  running on real hardware.''  This claim is extremely naive.  If a
  student writes an x86 program and runs it on a Pentium chip, is
  their software running on real hardware or via layers of firmware
  and emulation?  \emph{The answer depends on which model of chip
    is in their computer!}
\end{itemize}

\paragraph{Granularity of the time scale.}
The events that occur in a digital circuit take place on a short
time scale, with massive parallelism.  The events that occur at the
instruction set level exhibit far less parallelism, and the time
scale is two to three orders of magnitude longer.  These
differences in scale are interesting, and they do not cause
problems with simulation.  For example, experience shows that
students can follow all the details, and fully understand, the
processor circuit as it runs a real machine language program.
However, as we move up to the level of operating systems, the
varying time scales pose real difficulties.  To study the operation
of virtual memory in detail, we have to consider some events that
occur in a fraction of a clock cycle, and other events that occur
on the time scale of disk access.  These time scales differ by
\emph{nine orders of magnitude}.  It is challenging even to get
across to some students what the term ``nine orders of magnitude''
means.  However, these topics can be addressed in detail by
combining detailed circuit simulation where appropriate (e.g. the
TLB) with coarser grained emulation; the problem is that connecting
the dots between these levels must be done semiformally, with some
verbal explanation.

\paragraph{Models.}
A language isn't enough: it is important to describe clearly the
circuit models that are being used.  A model is an abstraction that
ignores some aspects of the hardware's behaviour, providing a view
that is simple enough to work with effectively.  The work described
in this paper is based on a pure synchronous model with a
single-phase clock.  Some other circuit models can be handled by
Hydra, but not all.  It is a fallacy to say that we don't need
models, because we will look at how the hardware ``really'' works.
``Real'' digital hardware actually consists of analogue circuits
that are carefully designed and operated so as to exhibit mostly
digital behaviour most of the time.  The details of this are
extraordinarily complicated \cite{W2000-DigDesign-Principles}, and
should not be discussed until and unless students have fully
mastered the synchronous model.

\paragraph{Karnaugh maps.}
Some students have an obsession with Karnaugh maps, which are
covered in the short section of the first year course on computer
hardware.  A Karnaugh map is an optimisation technique, not a
design technique, yet some students claim not to know how to design
a simple combinational circuit without starting with a Karnaugh
map.  There are two problems with this: (1) a basic principle in
computer science is to start with the specification, then perform
optimisation as needed; and (2) Karnaugh maps make a circuit more
efficient according to the component-count metric, but they may
make a circuit \emph{less} efficient according to other cost
models---especially the area-based cost models that are relevant to
VLSI design.  At best, a Karnaugh map is a small scale
transformation, akin to rearranging a few instructions to save an
instruction or two, but they do not address the larger scale
algorithmic issues that dominate circuit performance.

\subsection{Language issues}

\paragraph{Benefits of a functional hardware description language.}
The observed benefits include a good match between the foundation
of the language (functions) and the foundations of circuits
(functions); a precise language that allows clearer specifications
than are possible with vague diagrams; ability to define
sublanguages for clear description of specialised circuits such as
control algorithms; improved abstraction using design patterns
expressed as higher order functions; executable specifications;
typechecking circuit interfaces; and effective formal methods with
equational reasoning.

\paragraph{Language preferences.}
A number of students in the fourth-year course are extremely
enthusiastic about the functional hardware description language,
and are keen to use it further in projects, internships, or
research projects.  Feedback from these students indicates that
they like the power and simplicity of the specification, especially
the ability to design a digital circuit from the ground up that can
actually run computer programs---''connecting the dots''.  On the
other hand, some students in the fourth-year course say that VHDL
should be used because that is an industry standard language.  The
students who have given this feedback are doing a combined degree
with electrical engineering.

\paragraph{Prerequisites.}
No matter what language is used for describing hardware, there will
be some students who don't know it. To make the course materials
described in this paper more widely useful, an aim is to make the
course self-contained, so that knowledge of functional programming
in general or Haskell in particular is \emph{not} a prerequisite.
This is achieved by (1) not using the full Haskell language, but
rather a much smaller Hydra language; (2) teaching Hydra from the
ground up, which takes little time because the language is so
small; (3) implementing Hydra with a transformation tool that
provides an error message if the user steps outside Hydra into
Haskell.

\paragraph{Syntax.}
The semantics of a language is more fundamental than its surface
syntax, but some feel uncomfortable if the syntax looks
superficially unfamiliar.  The layout rule, in particular, seems to
help stronger students (the code is more concise and readable) but
to confuse weaker ones (who often avoid indentation entirely, and
just write each line of code beginning in the leftmost character
position).

\paragraph{Picky details.}
There are a lot of little details in digital circuits that can be
glossed over when giving a vague description, but that have to be
handled precisely correctly in order to get a correct specification
that can be simulated and executed successfully.  One example of
this is in bundling a group of signals into a cluster, such as a
word or tuple of signals.  Many publications on circuits, including
research papers as well as textbooks, use informal notations for
clusters of signals that give a general idea of what is going on,
but that rely on a full understanding in order to get the circuit
actually to work.  Hydra contains several kinds of machinery,
largely inherited from the Haskell type system, to help with these
details.  Nevertheless, it takes some time to cover the necessary
language features, and students often get type errors when they mix
up the signal clusters.  (These issues also arise in standard
languages like VHDL and informal notations such as used in the
Hennessey and Patterson books---if you want a precise
specification, you have to get the picky details right, and the
expressive type system of the functional language helps
significantly.)

\paragraph{Hardware or software?}
Any hardware description language will look like software to weaker
students, who will not understand the distinction between the
circuit and the software notations used to specify the circuit.
This happens because of inadequate thinking about abstraction, not
as a result of the choice of a functional or imperative hardware
description language.

\paragraph{DSL error messages.}
Hydra is a domain specific language (DSL) implemented by embedding in
Haskell.  The benefits of this approach have been discussed
extensively in the literature.  A drawback---which has also been
widely recognised---is that error messages often come from the host
language rather than from the DSL, so they make little sense to the
end user.  Thus it is possible to make an error in a circuit
specification, and to receive an error message that relates to Haskell
rather than to the circuit.  The new transformation system for Hydra
is currently being developed, and this appears to help greatly by
having an explicit representation of the Hydra language per se, but
further experimentation will be needed to evaluate this.

\subsection{Educational issues}

\paragraph{Expectations.}
It would be wonderful if the use of some secret magic bullet would
cause all students to master the computer systems material, and go
on to become real experts.  Alas, this just doesn't happen---not
even when functional programming is used.  The author's personal
experience is that the weakest students learn little, regardless of
what techniques are used.  However, the strongest students learn
much more using the functional hardware description language: they
really do achieve the working understanding which is the aim.

\paragraph{Exercises.}
The exercises, where students design circuits and observe their
execution, are vital to mastering the material.  It is tempting to
make all the exercises interesting and challenging, but that is a
mistake: It is essential to begin with some very straightforward
exercises to gain familiarity with the software tools.  It isn't
enough to give lots of examples, and then to assign an exercise
that assumes the students have assimilated the elementary aspects
of the examples.

\section{Conclusion}
\label{sec:conclusion}

It is better to use a hardware description language than just
showing some schematic diagrams when teaching circuit design,
especially for large and complex circuits.  A functional hardware
description language offers numerous additional benefits over an
imperative one: more concise specifications of circuits, a clear
connection to the view of circuits as functions, alternative
circuit semantics, and clean integration with formal methods.
Using this approach, it is possible to show a complete digital
circuit that fully implements a processor, with the ability to run
programs by simulating the circuit, and this greatly motivates the
stronger students.

\bibliographystyle{eptcs}
\bibliography{SysEduHydra}

\end{document}